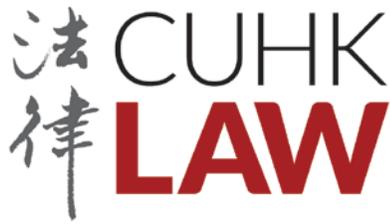

THE CHINESE UNIVERSITY OF HONG KONG

FACULTY OF LAW

Research Paper No. 2019-04

# LApps: Technological, Legal and Market Potentials of Blockchain Lightning Network Applications

Mahdi H. Miraz

David C. Donald

# LApps: Technological, Legal and Market Potentials of Blockchain Lightning Network Applications

Mahdi H. Miraz
The Chinese University of Hong Kong
Sha Tin
Hong Kong
m.miraz@ieee.org

David C. Donald
The Chinese University of Hong Kong
Sha Tin
Hong Kong
dcdonald@cuhk.edu.hk

## ABSTRACT
Following in the footsteps of pioneer Bitcoin, many altcoins as well as coloured coins have been being developed and merchandised adopting blockchain as the core enabling technology. However, since interoperability and scalability, due to high and capped (in particular cases) transaction latency are deep-rooted in the architecture of blockchain technology, they are by default inherited in any blockchain based applications. Lightning Network (LN) is one of the supporting technologies developed to eliminate this impediment of blockchain technology by facilitating instantaneous transfers of cryptos. Since the potentials of LN is still relatively unknown, this paper investigates the current states of development along with possible non-monetary usage of LN, especially in settlement coloured coins such as securities, as well as creation of new business models based on Lightning Applications (LApps) and microchannel payments as well as micro-trades. The legal challenges that may act as impediment to the adoption of LN is also discussed.

## 1. INTRODUCTION
Despite several successful new and experimental applications of blockchain [1] technology, including the conceiving Bitcoin, there are few inbred limitations that caveat the outright utilisation of the technology. Lack of interoperability amongst various chains and restrained scalability by high transaction latency due to decentralised consensus approach and network architectural limitations are the prominent ones [2]. For instance, hitherto Bitcoin can process 7 transactions per second (TPS) and Ethereum has the capacity of 15 TPS. While Ripple has a capacity of 1500 TPS, it is far lower than that of Visa which is 24,000 [3] and if demand increases it is likely to go down due to increased network congestions and augmented load on the consensus process. At least, this has been the case for Bitcoin and Ethereum: An "unconfirmed" Bitcoin transaction might take a minimum of 10 minutes on an average to get "confirmed" if pulled into the next available "candidate" block, otherwise the process can take up to several days. Due to increasing popularity of DApps (Decentralised Apps), DAOs (Decentralised Autonomous Organization) and ICO's (Initial Coin Offering), the waiting time in Ethereum network is also exponentially rising. Off-chain scaling, such as lightning network (LN) [4,5], can play a vital role in this regard. In off-chain scaling [4,6,7,8,9,10,11], a second layer, also known as layer 2 or payment channel, enables unlimited instantaneous transactions to take place between two parties. When the channel is closed, only the netted result of the transactions is broadcast to the network for consensus. LN thus directly holds the promises of eliminating the scalability problem of blockchain while atomic swaps powered by a LN possess the potential of enabling smoother interoperability.

The rest of the paper briefly describes the fundamental principle that serves as the foundation for LN. The paper then evaluates the feasibility of its application in exchanging and settling of coloured coins – real world assets represented on the Bitcoin (or more widely any blockchain) networks such as securities, bonds, futures, shares and other commodities. The paper also discusses the creation of new non-monetary business models laid on the foundation of LN – mainly utilising Lightning Applications (LApps) and LN empowered micropayment provisions.

## 2. FUNDAMENTALS OF LIGHTNING NETWORK
### 2.1 Preamble
A Lightning Network is a second layer Hashed Timelock Contract (HTLC) based smart contract enabling bi-directional payment channels built on top of a base layer of blockchain such as Bitcoin. LN effectuates secure routing of payments or transactions of coloured coins across multiple peer-to-peer (P2P) payment channels enabling transactions between two parties who are not directly connected by any point-to-point channel. Thus, by off-loading the transactions away from the base layer, LN engineers instantaneous transfers of assets, cryptocurrencies or other crypto assets with near-zero transaction fees. The concept of lightning network was first revealed in December 2015 by Joseph Poon and Thaddeus Dryja [4], however, it took almost two years to undertake a series of successful implementations of interoperable test transactions on Bitcoin core network in December 2017.

Let us consider that Alice and Bob would like to establish a LN payment channel for transacting Bitcoins (BTCs) amongst themselves. The first step is creating a multisignature wallet which can be accessed by both of them using their respective private keys.

After the wallets is created, they then need to make a deposit, the "funding transaction", a certain amount of BTC (for example 10 BTC) each into the multisignature wallet they have created. A "commitment transaction" is then required to enable the transfer of funds using LN. For instance, Alice wants to send 2 BTC to Bob, she will simply need to transfer ownership of 2 BTC to bob having the new balance sheet (Alice owns 8 BTC while Bob now owns 12 BTC) signed by the private respective private keys of both parties. They can now conduct unlimited transactions between them by redistributing the funds of the shared wallet. The transfer of ownership rights is bi-directional and can be performed an unlimited number of times before the channel is closed by either party. In this scenario, the funds are actually distributed at the time of closing the channel and initial as well as final balance are then broadcasted to the peers of the base blockchain for consensus approach as normally performed in any base layer transactions. Thus, the outcome of netted multiple LN transactions is recorded on the blockchain as a single transaction.

In fact, the commitment transaction fundamentally allocates the funding transactions as per the current allocation and therefore comprises two asymmetric transactions. In case of Alice, these are: one that pays Bob straightaway and the other that is a revocable but time-locked output, which ultimately pays Alice. The later can be revoked by Bob if he knows the revocation key. Similarly, Bob's commitment transaction will be the converse.

Let us consider the case where Alice is connected with Bob using one LN payment channel while Bob is connected to Trudy by another channel. This will enable Alice to indirectly transfer her coins to Trudy via Bob without needing any dedicated channel between them. With the widespread adoption of LN, such indirect channel will automatically increase in scope. However, a routing algorithm will be required to find an optimal route from the source to the destination. LN adopts an onion style routing approach without comprising the privacy where the intermediate nodes only know the next hop address rather than both the next hop and the final destination addresses as in traditional routing algorithms. As of 15:06:40 GMT+0800 (Hong Kong Standard Time) on Thursday 31$^{st}$ January 2019, the number of total number of "reachable" Bitcoin nodes was 10,527 with an average of 10,301 in the last 24 hours[1], while the number of LN enabled nodes was 5,788 with 23,021 channels of a total network capacity (for LN transfers) of 618.51 BTC. However, the number of "active" LN enabled nodes was only 2,870[2].

So far, there have been several variant implementations of the originally proposed LN, following recommendations from other developers of the Bitcoin community. The three major implementations are: Blockstream's "c-lightning" implementation in C, Lightning Labs' Golang's implementation of "Lightning Network Daemon (LND)" and ACINQ's Scala implementation of "eclair". A complete updated list can be found at GitHub.[3] All three of these have been proven interoperable by real LN transfers. Ethereum's Raiden Network is also an example of off-chain scaling similar to LN.

### 2.2 Basic Algorithm
Initially, Alice's commitment transaction is A1 with a revocation key of RA1 which is only known by Alice. Similarly, Bob's commitment transaction is B1 with a revocation key of RB1 which is only known by Bob.

Let us assume Alice wants to send Bob 2 BTC (initially she had 10 BTC).

1. Alice generates a new Bob's transaction B2 allocating 8 BTC to Alice and 12 BTC to Bob.
2. Alice then signs B2 with her private key and transmit it to Bob.
3. Once received, Bob signs B2 and temporarily keeps it
4. Bob generates a new Alice's transaction A2 allocating 8 BTC to Alice and 12 BTC to Bob.
5. Bob then signs A2 with his private key and transmit it to Alice.
6. Once received, Alice signs A2 and temporarily keeps it
7. Alice shares RA1 invalidating A1; A1 can now be deleted
8. Bob shares RB1 invalidating B1; B1 can now be deleted

**Algorithm 1: Basic LN Algorithm, developed based on the original LN proposal by Poon and Dryja** [4]**.**

### 2.3 Major Advantages
Although the primary intension of developing LN was to facilitate instantaneous payments over Bitcoin networks, it brings many other advantages such as:

- Since LN is the enabler of off-chain atomic cross-chain swaps [2], all the benefits atomic swap can offer are imputed to LN including those of sidechains.
- Off-chain scaling such as LN will help cryptocurrencies to compete with fiat currencies to some extent.
- Off-loading some transactions away from the base layer of chain will shorten the processing queue of "unconfirmed" transactions which will result in reduced on-chain transaction fees.
- Improved privacy is another key advantage of LN as the transactions are not recorded on the base DLT. Onion style nested routing approach adds an extra layer of privacy as the intermediate hops can only see the next hop's address, without revealing the final destination address.
- Merchants of commodities such as online shops or food outlets can open a LN Channel and receive instant payments.
- Since the transaction fee is near-zero, LN effectively works as a micropayment channel.
- LN based LApps possess great potentials to lead the creation of new ventures and innovative business models.
- LN actuate micro-trading of cryptocurrencies and other crypto assets.
- LN has the potentials of being used in the settlement of non-monetary coloured coins such as securities.

### 2.4 Major Impediments
Considering the fact that off-chain scaling technology such as LN is still in its infancy, there are many impediments to overcome before it is widely and effectively adopted. Thus far, the major constrains of this technology are as follows:

- LN is considered to be a "resource hog" since both the Bitcoin and LN nodes need to be run on the same server. This requires extremely high computational power as well as considerable amount of time for new installations, especially due to the time required for synchronisation of the blockchain, currently sized at 196.56 GB.[4]
- Although LN enabled transactions are instantaneous compared to on-chain transactions, it is still very slow compared to fiat payment systems such as Visa or Master.
- If at any time either party drops or goes offline, the channel will be closed and settled.
- Off-chain scaling is not yet supported in many altcoins.
- LN is created as a separate layer (layer 2) on top of the base blockchain layer; therefore, it doesn't inherit the security features of blockchain. Considering the fact that the technology is still not highly proved to be secure, the supporting networks limit the amount of the currency to be traded.
- Crypto systems without smart contract support cannot facilitate off-chain scaling.
- Implementing off-chain scaling requires extensive programming skill.
- LN, in its current form, is highly vulnerable to distributed denial of service (DDoS) [12] and other cyber-attacks [13]. In fact, LN already faced a DDoS on the 20 March 2018 that sent approximately 200 nodes offline, which is roughly 20% of the total available LN nodes prior to the attack.
- LN is not light-weight and is highly interdependent on complex technological configuration. Configuring cluster of servers, as seen in traditional e-commerce or in banking for redundancy, is very complex in the current design of LN. Therefore, LN for business application are not redundant and susceptible to single point of failure (SPF).
- Implementing LN in any dynamic cloud environment will demand significant workaround of the current design.

---

[1] https://bitnodes.earn.com/dashboard/

[2] https://1ml.com/statistics

[3] https://github.com/bcongdon/awesome-lightning-network

# 3. APPLICATIONS OF LIGHTING NETWORK IN NON-MONETARY TRANSACTIONS AND FUTURE ADOPTION TRENDS

Lightning network technology is still a new concept-provisionally developed and implemented with limited scope. The number and capacity of lightning transactions taking place at this moment is still very small. However, introduction of LN is gaining popularity as it solves some of the problems associated with blockchain technologies, in particular with Bitcoin blockchain.

Currently there are only a few crypto systems that support both HTLC and the specialised programming functions which are the minimum technological requirements to adopt LN. However, they are expected to implement these features in the future, which will highly determine the direction of LN's adoption trends.

Rauchs *et al.* [14] describes LN as an example of an "interfacing" Digital Ledger Technology (DLT) system that "opportunistically" implements the elemental functionalities provided by a DLT technical configuration, which could be adjusted to take advantage of another or even multiple base DLTs. This feature of LN not only enables off-chain atomic cross-chain swaps [2], but also materialises the notion of decentralised exchanges, as each LN node having channels linked to multiple blockchain networks can act as a decentralised exchange.

Since nodes in the LN act as a money transmitter, there is a debate whether registration as a money transfer is a legal requirement. However, the laws related to money transmission differ across various countries, even sometimes among states of a country, such as in the US. Laws of many countries or states even lack clarity on whether the law for fiat currencies can be applied to cryptos. In fact, the nodes of LN do not acquire real ownership of the funds while being transmitted. Therefore, they cannot possibly nobble. Effectiveness of applying money transfer law in this case thus may not justify its intention to protect consumers.

In fact, such inter-blockchain interoperability, i.e. atomic swaps will help boost liquidity of crypto assets arising in these chains by enabling transfer of assets formulated on one chain into assets formulated on another chain. Such swaps could include settlement of securities transactions.

For legal certainty, the current governance approach of standard framework agreements applied in international swap transactions can also be used for atomic swaps, with very little adjustment necessary once the framework is set in place.

From a regulatory viewpoint, swaps and/or transfers present the problem that they could traverse geographic boundaries of political entities or legal jurisdictions. Such transnational activity is harder to regulate and monitor by the regulatory bodies or similar government agencies of any jurisdiction. Secrecy is added to this regulatory difficulty. Due to the implantation of an onion style routing approach, even if LN channels created by other users are used to facilitate a transaction or swap, and only the final netted balance is broadcasted to the base blockchain network for consensus, the intermediate transfers or swaps remain private. No one but the transacting parties knows the actual transaction details. This feature of LN can contribute to the rise of illicit markets.

LN-powered atomic swaps have the potentials of eliminating legacy crypto exchanges by deploying direct transfers as well as decentralised exchanges. This could make monitoring and regulation more difficult. Since the intermediate transactions are not recorded even by the transacting parties, the inspection of chain coding is less likely to be sufficient. Therefore, the future adoption models of crypto assets is highly dependent on how regulatory and legal provisions are adjusted in different jurisdictions.

In parallel with the pervasive application of blockchain technologies [15,16], LN even holds the promise to widen the current scope of adoption, especially LN could potentially play a role in settlement of securities transfers [17] or of other coloured coins - bringing the direct and transparent holding of assets back to organised markets [18]. LN could thus contribute to higher transparency leading to better corporate governance.

LN enabled off-chain atomic cross-chain swaps [2] could also play an important role in facilitating cross-listing on blockchain based securities exchanges. As per the current cross-listing models, one company can list its security in multiple exchanges [19]; the emerging concept of sidechain could be adapted for this purpose. A sidechain, also known as childchain in Ardor platform, is a 'loosely' joined independent blockchain, attached to another (i.e.

---

[4] https://www.blockchain.com/en/charts/blocks-size

parent) blockchain utilising the "two-way peg" approach. This enables crypto assets from the parent blockchain to be securely moved and used in the sidechains, at an agreed rate, with the option to move back to the original (i.e. parent) chain.

Analogous to making micro-payments for any commodity, Micro-trades of cryptocurrencies (as in Foreign Exchange) or other coloured coins such as shares and securities, can also be feasible using LN.

Despite the development of the various LN testbeds, seamless multi-asset conversion is still limited by its technical design. Therefore, adoption of such trade is foreseeable but subject to the maturity of the technology.

In terms of using LN for American Call Options or for any other scenarios where significant changes in the traded value of the assets may happen within the trade windows, with current technological limitations of HTLCs, it is still problematic. This is primarily because traders can take unfair advantage of the current contracts if a rate changes in their favour. However, if LN aims to handle such trades, either carefully designed modification of the technology or development of a legal mechanism will be required [20].

As the LN technology has existed in operable form for just around one year and is still being used in restricted manner, we consider this still to be a research and development (R&D) phase. However, as LN technology matures, its more concrete utilisation of the technology in various applications is expected, especially for transfer and fungibility of digital assets.

## 4. DRIVERS OF FUTURE BUSINESS MODELS

Among all the benefits of LN as described in section 2.2, following two are currently the enablers of new business models and innovation:

1. Lightning Applications (Lapps) and
2. Micropayment Channel

### 4.1 Lightning Applications (LApps)

One of the major drawbacks of Bitcoin, compared to other Blockchain 2.0 implementation such as Ethereum, is not being "turing-complete" and thus having "no" support for smart contracts. However, implementation of off-chain scaling via LN Bitcoin blockchain can now support lightning applications (LApps) – similar to decentralised applications (DApps). This is made possible primarily due to off-chain transactions and multisignature features of LN. Combined with near-zero transaction fees, micropayment facilities and instant fund transfer, many business use-cases are being developed.

### 4.2 Micropayment Channel

A micropayment, as the name implies, facilitates financial transactions of a very small amount, usually using online facilities. In the 1990s, many micropayment channels were implemented but could not gain mass popularity, mostly due to limited use of the Internet. Utilising emerging e-commerce technologies, the concept of micropayment again came into highlight. In 2010s, the ecommerce industry has seen a second generation of micropayment. In fact, the success of a micropayment system highly depends on the capability of offering extremely low transaction fees. Since LN can offer near-zero transaction fees, it is considered to be an enabler of micropayment channels. To help promote micropayment by LN, a tiny artwork named "Black Swan" has been auctioned and sold by a user to the lowest bidder for 1 milli-satoshi i.e. $0.000000037 - the lowest possible amount that could be transferred by Bitcoin LN[5]. If not the lowest, this is certainly close to being the lowest price ever paid in an auction.

### 4.3 Applications and Adoption Trends of Lightning Network in E-Commerce

Until LN was implemented, the use-cases for Bitcoin end users was extremely limited to financial activities – mainly managing and funding wallets as well as exchanges. LN widens the scope and holds promises to further create new business models empowered by instant payments, even at micro level with near-zero transaction fees. This will thus help significantly lower threshold of entry barriers for new businesses. Examples of such implementations include platforms for monetising digital contents at micro level, international mobile messaging services, in-game micro-payments, receiving tips, Lightning Jukebox, Gambling, Lightning Point-of-Sale (PoS) for non-virtual stores, Bitcoin-payable Twitter Bot that facilitates receiving payments for like, share (retweets) and follows are few to mention. Lightning App Directory provides a comprehensive list of applications based on LN.[6]

## 5. CONCLUDING REMARKS

By raising the limitations inherent in the two major failures of any blockchain-based financial application, interoperability and scalability, this paper argues that a second layer solution such as lightning network, powered by HTLC and smart contract, has potential for addressing these limitations. The paper then discusses prospective usage of lighting networks in the clearing and settlement of coloured coins and securities. The paper also discusses future adoption trends of lightning networks and thus the creation of new business models utilising this technology. Legal and regulatory aspects of LN are discussed and future research directions projected.

---

[5] http://bitcoinist.com/lightning-network-black-swan-cryptograffiti

[6] https://dev.lightning.community/lapps/